\documentstyle[12pt,graphicx,subfigure]{article}
\def\slash#1{\setbox0=\hbox{$#1$}#1\hskip-\wd0\hbox to\wd0{\hss\sl/\/\hss}}
\begin{document}
\baselineskip=20 pt
\def\l{\lambda}
\def\L{\Lambda}
\def\b{\beta}
\def\a{\alpha}
\def\d{\delta}
\def\g{\gamma}
\def\mphi{m_{\phi}}
\def\dnul{\partial_{\nu}}
\def\dnuu{\partial^{\nu}}
\def\dmul{\partial_{\mu}}
\def\dmuu{\partial^{\mu}}
\def\eps{\epsilon}
\def\hphi{\hat{\phi}}
\def\vphi{\langle \phi \rangle}
\def\mph{m_\phi}
\def\etamunu{\eta^{\mu\nu}}
\def\bfl{\begin{flushleft}}
\def\efl{\end{flushleft}}
\def\bea{\begin{eqnarray}}
\def\eea{\end{eqnarray}}
\def\la{\langle}
\def\ra{\rangle}
\begin{center}
{\large\bf {Muon anomaly and a lower bound on higgs mass 
due to a light
stabilized radion in the Randall-Sundrum model.\\}}  
\end{center}
\vskip 5pT
\begin{center}
{\large\sl {Prasanta Kumar Das}~\footnote{Email:pdas@mri.ernet.in,~dasp@imsc.res.in} 
}
\vskip  5pT
{\rm
Harish-Chandra Research Institute, \\
Chhatnag Road, Jhusi,\\
Allahabad-211019, India .}\\
\vskip  5pT
{\rm
{\it presently at} \\
Institute of Mathematical Sciences \\
C.I.T Campus, Taramani, \\
Chennai-600113, India}\\

\end{center}

\vskip 10pT
\centerline{\bf Abstract}

\vskip 5pT
\noindent 
{\it We investigate the Randall-Sundrum model with a light stabilized 
radion (required to fix 
the size of the extra dimension) in the light of muon anomalous magnetic 
moment $a_\mu [= \frac{(g - 2)}{2}]$. Using the recent data (obtained from the E821 
experiment of the 
BNL collaboration) which differs by $2.6~\sigma$ from the Standard Model result, 
we obtain constraints on radion mass $\mphi$ and radion vev $\vphi$. 
In the presence of a radion the beta functions $\beta(\l)$ and $\beta(g_t)$ of higgs 
quartic 
coupling ($\l$) and top-Yukawa coupling  ($g_t$) gets modified. We find these modified beta functions. 
Using these beta functions together with the 
anomaly constrained $\mphi$ and $\vphi$, we obtain lower bound on higgs mass $m_h$.  We 
compare  
our result with the present LEP2 bound on $m_h$. }

\bfl
{\it Keywords}: Extra dimensional field theory; Renormalization; Higgs boson.\\
{\it PACS Nos.}: 11.10.Kk, 11.10.Gh, 14.80.Bn
\efl

\vskip 5pT
\newpage

\section{Introduction}
Recently the notion of extra spatial dimension(s)
\cite{ADD1,ADD2,RS1,RS2}, proposed as a resolution of the hierarchy problem, draws a lot of attention 
among the physics community. Among these the 
Randall-Sundrum(RS) model (of warped spatial dimension) is particularly interesting from the 
phenomenological point of view\cite{GMPK,MR,PS,KimKim}. This model views the world 
as $5$-dimensional and it's fifth spatial dimension is 
$S^1/Z_2$ orbifold. The metric of such a world can be written as
\bea
d s^2 = \Omega^2 \eta_{\mu \nu} d x^\mu d x^\nu
- R_c^2 d \theta^2.
\eea
\noindent
The factor $\Omega^2 = e^{-2 k R_c |\theta|}$  is called the warp factor. In $\Omega^2$,   
$k$ stands for the bulk curvature constant and $R_c$ corresponds to the size
of the extra dimension. The angular variable $\theta$ parametrizes the fifth 
dimension. The model is constructed out of  
two $D_3$ branes located at the orbifold fixed points. 
$\theta = 0$ and $\theta = \pi$ respectively. The brane located at  
$\theta = 0$ (where gravity peaks) is known as the Planck brane, while that located at 
$\theta = \pi$ (the SM fields resides on it and the gravity is weak) is called the TeV 
brane. 

The radius $R_c$(distance between two branes) can be 
related to the vacuum expectation value (vev) of some modulus
field $T(x)$ which corresponds to the fluctuations of the metric over the
background geometry given by $R_c$. Replacing $R_c$ by $T(x)$ we can rewrite
the RS metric at the orbifold point $\theta = \pi$ as
\bea
d s^2 = g_{\mu \nu}^{vis} d x^\mu d x^\nu, 
\eea

\noindent
where $g_{\mu \nu}^{vis} = e^{- 2 \pi k T(x)}\eta_{\mu \nu}
= \left(\frac{\phi(x)}{f}\right)^2 \eta_{\mu \nu}$. Here
$f^2 = \frac{24 M_5^3}{K}$, $M_5$ is the $5$-dimensional Planck scale
\cite{RS1,RS2} and $\phi(x) = f e^{- \pi k T(x)}$. The scalar field $\hat{\phi}(x)$ 
(i.e. $\hat{\phi}(x) = \phi(x) - \vphi $) is known as the radion 
field \cite{GRW,GW,GR}. In the minimal version of the RS model there is no potential which can stabilize the 
modulus field $T(x)$ (and thus the radion $\hat{\phi}(x)$). However in a pioneering work Goldberger and 
Wise \cite{GW1,GW2} were able to generate a potential of this modulus field (by adding an extra massive bulk 
scalar) field which has the correct minima satisfying $k R_{c} \simeq 11$, a necessary 
condition for the hierarchy resolution. 

In this non-minimal RS model (RS model together with the Goldberger and Wise mechanism), the 
stabilzed radion can be lighter than the 
other low-lying gravitonic degrees of freedom and will reveal  
itself first either in the direct collider search 
or indirectly through the precission measurement. Studies based on observable consequences 
of radion are available in the literature \cite{Cheung,Csaki,UmaDat,BKL,DM1,DM2,CDHY}. Here we 
will make one such study in the light of muon anomalous magnetic moment
$a_\mu = \frac{(g - 2)}{2}$. 

 The E821 experiment \cite{E821ex1,E821ex2} of the BNL collaboration recently has reported
a new measurement of muon magnetic moment ($a_\mu^{(expt)}$) which is a positive one
and deviates from the SM based calculation by $2.6~\sigma$. The measured experimental value  
$a_\mu^{(exp)}$ lies in the range
\bea \label{eqn:muonanlyBNL}
a_\mu^{(expt)} = (11659204(7)(4)) \times 10^{-10}
\eea 
in units of Bohr magneton $e/{2m_\mu}$. Comparing this and the present 
SM result (which includes QED, electroweak
and hadronic contribution) \cite{Hert,KN,KNPR,HayKin,BCM} which is about 
\bea \label{eqn:muonanlySM}
a_\mu^{(SM)} = (11659176 \pm 6.7) \times 10^{-10},
\eea
\noindent one finds a lot of  option to explain the extra contribution
$\delta a_\mu^{new} [= a_\mu^{(expt)} - a_\mu^{(SM)}]$ by means of some  
non-standard new physics \cite{CzarMar,Lane,EKRW,FenMat,BalGon,UtpalNath,Mahanta,CCD,CMR,BarMai,KKS,ACN,LNW,AEW,ChatNath,CGW,NathYama,DHR,MarCzar}. 
At the same time by taking a conservative viewpoint one can also constrain the new physics by  using 
this $\delta a_\mu^{new}$.

The organization of the paper is as follows. In section 2, we describe the effective 
interaction of radion and the SM fields (on the TeV brane). We  
obtain the effective renormalized higgs quartic and top-Yukawa 
couplings $\l$ and 
$g_t$ in the presence of radion and find the corresponding modified beta functions 
$\beta(\l)$ and $\beta(g_t)$. We find the radion contribution to muon anomalous 
magnetic moment $a^\phi_\mu$.  Section 3 is devoted for the numerical
analysis.  Defining the excess $\delta a_\mu^{new} [= a_\mu^{(expt)} - 
a_\mu^{(SM)}]$ in terms of $a^\phi_\mu$ (the radion contribution), we obtain constraints on 
radion mass $m_\phi$ and radion vev 
$\vphi$ which are the two free parameters of this model. 
Using these anomaly constrained $m_\phi$ and $\vphi$ and the modified beta functions 
$\beta(\l)$ and $\beta(g_t)$ we then obtain lower bounds on the
SM higgs mass $m_h$. We compare our result with the LEP2 
bound on $m_h$ obtained from the direct search \cite{Junk}. Finally we summarize our results and 
conclude.  

\vspace*{-0.2in}

\section{Effective Interactions and Renormalization}
Radion interaction with the SM fields(residing on the TeV brane) is governed by the 
$4$ dimensional general coordinate invariance. It couples to the trace of the
energy-momentum tensor of the SM fields as 
\bea \label{eqn:radcoup}
{\mathcal{L}}_{\it int} = \frac{\hat{\phi}}{\vphi} T^\mu_\mu (SM),
\eea
where $\vphi$ is the radion vev and $T^\mu_\mu (SM)$ is given by 
\bea
T^\mu_\mu (SM) = \sum_{\psi} \left[\frac{3 i}{2} \left({\overline{\psi}}
\g_\mu \dnul \psi - \dnul{\overline{\psi}} \g_\mu \psi \right)\eta^{\mu\nu}
- 4 m_\psi {\overline{\psi}} \psi\right] - 2 m_W^2 W_\mu^+ W^{-\mu}
- m_Z^2 Z_\mu Z^\mu \nonumber \\
+ (2 m_h^2 h^2 - \partial_\mu h \partial^\mu h) + ...
\eea
Note that in order to accommodate the gauge interaction(s)  with fermion-radion system one has to write 
the ordinary derivative to a gauge covariant derivative.
The photon and the gluons couple to the radion
via the usual top-loop diagrams\cite{DRR}. Besides this there is an added source of enhancement
of the coupling due to the trace anomaly term (See \cite{CDJ} for a nice discussion) which is given by
\bea
T^\mu_\mu (SM)^{anom} = \sum_{a} \frac{\b_a(g_a)}{2 g_a} G_{\mu\nu}^a
G^{a \mu\nu}.
\eea
For gluons $\b_s (g_s)/{2 g_s} = - [\a_s/{8 \pi}]~ b_{QCD}$ where
$b_{QCD} = 11 - 2 n_f/3$ and $n_f$ is the number of quark flavours. 
In the next subsection we derive in detail the radion interaction with the 
higgs scalar and the top quark as they will be relevant in our course of finding 
renormalized $\l$ and $g_t$.

\subsection{Radion-higgs coupling and $\l$ renormalization }
The radion coupling to the higgs scalar is completely determined by
$4$ dimensional general covariance. The action for the higgs scalar in the 
Randall-Sundrum model can be written as
\bea
S=\int d^4x \sqrt {-g_{vis}}[g_{vis}^{\mu\nu}{1\over 2}\dmul h \dnul h -V(h)],
\eea
where $V(h)={1\over 2} \mu^2 h^2 +{\l \over 4} h^4$. h is a small
fluctuation of the higgs field from its classical vacuum v. 
In absence of graviton fluctuations we have
$g_{vis}^{\mu\nu}= e^{2k\pi T(x)}\eta^{\mu\nu}=\left({\phi\over f}\right)^{-2}
{\eta^{\mu\nu}}$ and $\sqrt{-g_{vis}} =\left({\phi\over f}\right)^4 $ where 
$\phi = f e^{-k\pi T(x)} = \langle \phi \rangle + \hat{\phi}$.  Rescaling h
and v as $h\rightarrow {f\over \vphi}h$ and $v\rightarrow
{f\over \vphi} v$ we get
\bea
S=\int d^{4}x [(1 + \frac{\hat{\phi}}{\vphi})^2{1\over 2}{\eta^{\mu\nu}}\partial_{\mu} h \partial_{\nu} h-
(1 + \frac{\hat{\phi}}{\vphi})^4 V(h)],
\eea
where $V(h)={\lambda\over 4} (h^4+4h^3v+4h^2v^2)$.

The Feynman diagrams that give rise to the radion (in figures $\phi \equiv \hat{\phi}$) contribution to 
the renormalization of the four higgs vertex  in the RS model are
shown in Figure 1a. 


\begin{figure}[htb]
\begin{center}
\vspace*{1.2in}
      \relax\noindent\hskip -6.4in\relax{\includegraphics{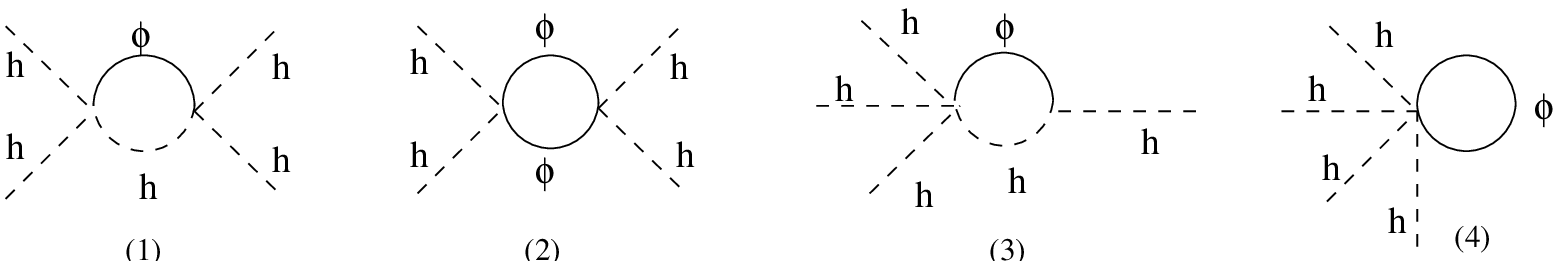}}
\end{center}
\end{figure}
\vspace*{0.05in}
\noindent {Figure 1a:}
{\small { Feynman diagrams that give rise to the radion contribution
  to the vertex renormalization . Here $\phi$ corresponds to $\hat {\phi}$.}} 

We now evaluate the vertex renormalization diagrams explicitly with a cut
off $\L$ and the leading log terms of these diagrams are given by
\bea
\Gamma_1 = 6\l {288\l v^2\over 16\pi^2\vphi^2}\ln {\L^2\over \mu^2},
\eea
\bea
\Gamma_2 =6\l {144\l v^4 \over 16\pi^2 \vphi^4} \ln{\L^2\over \mu^2},
\eea
\bea
\Gamma_3 = 6\l {128\l v^2\over 16\pi^2\vphi^2}\ln {\L^2\over \mu^2},
\eea
and
\bea
\Gamma_4 = -6\l {6\over 16\pi^2\vphi^2} [\L^2 -\mphi^2 
\ln {\L^2\over \mu^2}].
\eea
Here $\mu$ is the renormalization scale. In the SM model the 
wavefunction renormalization constant of the higgs boson $Z_h$
is equal to one at one loop order even after the higgs field is 
shifted by its vev. However the radion coupling to the KE term 
of the higgs boson gives rise to a non-trivial wavefunction
renormalization of the higgs boson. Evaluating the radion mediated
self energy diagram (Figure 1b) of the higgs boson and considering the terms proportional to $p^2$   
we find that
$Z_h= 1+{1\over 32\pi^2}{5 m_h^2-\mphi^2 \over \vphi^2}\ln {\L^2\over
\mu^2}$.

\vspace*{0.4in}
\begin{figure}[htb]
\begin{center}
\vspace*{0.5in}
      \relax\noindent\hskip -3.2in\relax{\includegraphics{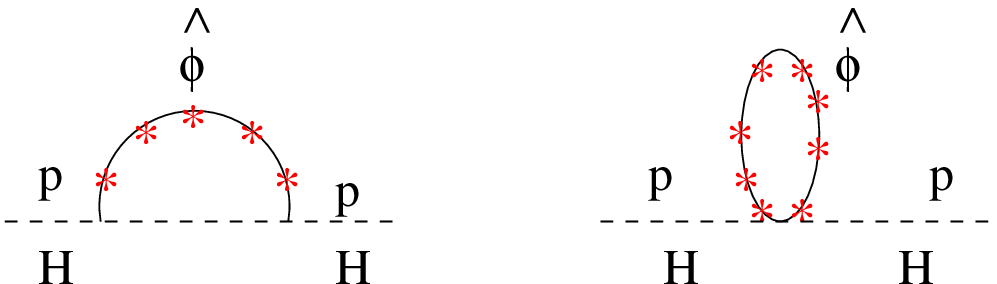}}
\end{center}
\end{figure}
\vspace*{-0.35in}
\noindent {Figure 1b:}
{\small { Radion mediated self-energy diagram of the higgs boson.}} 

\vskip 10pt

Using the above vertex and wavefunction renormalizations induced by
a light radion it can be shown that the complete one loop beta
function for $\l$ in the RS model is given by \cite{DasUma1}
\bea \label{eqn:hdaRad}
{\beta (\lambda )}  = \mu {d\lambda \over d\mu }={1\over 8\pi^2}[9\lambda^2 +
{402 \lambda^2 v^2\over \vphi^2}+ {144\lambda^2 v^4\over \vphi^4} + {5\lambda \mphi^2
\over \vphi^2}
+\lambda (6 g_y^2-{9\over 2} g^2-{3\over 2}g^{\prime 2})] 
\nonumber \\
+\frac{1}{8 \pi^2} [-6 g_y^4+{3\over 16} (g^4+{1\over 2}(g^2+g^{\prime 2})^2)].
\eea

\noindent The purely SM contribution to $\beta(\lambda)$ (see the next subsection)
can be obtained by letting the expansion parameter $\vphi$ approach infinity.

\subsection{Radion-top coupling and $g_t$ renormalization }
The radion coupling to the top quark in the Randall-Sundrum model
can be derived from the following action
\bea
S_1 = \int d^4 x \sqrt{- g_v} \left[{\overline{\psi}}\left(i
\gamma_a e^{a\mu}D_\mu - m \right)\psi - \frac{g_t}{\sqrt{2}} H
{\overline{\psi}} \psi \right], 
\eea
where $e^{a\mu}$ is the contravariant vierbein field for the visible
brane. In the presence of radion fluctuation it satifies the normalization
condition 
\bea
e^{a\mu}e_{a}^{\nu} = g^{\mu\nu} =
\left(\frac{\phi}{f}\right)^{-2}~\eta^{\mu\nu} =
e^{2 \pi k T(x)}~\eta^{\mu\nu}.
\eea
$D_\mu$ is the covariant derivative with respect to general coordinate
transformation and is given by 
\bea
D_\mu \psi = \partial_\mu \psi + \frac{1}{2}w_{\mu}^{ab} \Sigma_{ab} \psi.
\eea 
Here $\Sigma_{ab} = \frac{1}{4} \left[\gamma_a,\gamma_b \right]$.
The spin connection $w_{\mu}^{ab}$ in terms of the vierbein fields reads as
\bea
w_{\mu}^{ab} = \frac{1}{2} e^{\nu a}(\partial_\mu e^b_\nu - \partial_\nu
e^b_\mu) - \frac{1}{2} e^{\nu b}(\partial_\mu e^a_\nu -
\partial_\nu e^a_\mu) -  \frac{1}{2} e^{\rho a}e^{\sigma b}(\partial_\rho
e_{\sigma c} - \partial_\sigma e_{\rho c})e^c_\mu. 
\eea
It can be shown that in the presence of
radion fluctuations on the visible brane the spin connection is given by
\bea
w_{\mu}^{ab} = \frac{1}{\phi} \partial_\nu \phi \left[e^{\nu
b}e^a_\mu - e^{\nu a} e^b_\mu\right].
\eea
The covariant derivative of the fermion field becomes 
\bea
D_\mu \psi = \partial_\mu \psi + \frac{1}{4
\phi}\partial^\nu \phi \left[\gamma_\mu,\gamma_\nu\right] \psi,
\eea
where the $\gamma_\mu$ are independent of space time coordinates. The
action comprising radion coupling to the top quark finally can be
written as
\bea \label{eqn:radtop}
S_1 = \int d^4 x \left(\frac{\phi}{f}\right)^4
\left[\left(\frac{\phi}{f}\right)^{-1} {\overline{\psi}}\{i
\gamma^\mu \partial_\mu + \frac{3 i}{2 \phi} \partial_\mu \phi
\gamma^\mu \}\psi - m_t {\overline{\psi}} \psi -
\frac{g_t}{\sqrt{2}} H {\overline{\psi}} \psi \right] \noindent
\nonumber \\
= \int d^4 x \left[{\overline{\tilde{\psi}}}\{i \gamma^\mu \partial_\mu
{\tilde \psi} + \frac{3 i}{2 \phi}\partial_\mu \phi \gamma^\mu
{\tilde \psi} \}\left(1 + \frac{\hat{\phi}}{\vphi}\right)^3 -
\left(\tilde{m_t} + \frac{g_t}{\sqrt{2}} \tilde{H}\right)\left(1 +
\frac{\hat{\phi}}{\vphi}\right)^4 {\overline{\tilde{\psi}}} {\tilde \psi}
\right] \noindent
\nonumber \\
= \int d^4 x \left[{\overline{\tilde{\psi}}}i \gamma^\mu \partial_\mu
{\tilde \psi} - \tilde{m_t} {\overline{\tilde{\psi}}} {\tilde{\psi}} -
\frac{g_t}{\sqrt{2}} \tilde{H} {\overline{\tilde{\psi}}} {\tilde
\psi}\right] \noindent
\nonumber \\
+  \int d^4 x \left[ \frac{3 i}{ \vphi} {\overline{\tilde{\psi}}}
\gamma^\mu \partial_\mu {\tilde \psi}~ {\hat{\phi}} + \frac{3 i}{2 \vphi}
{\overline{\tilde{\psi}}} \gamma^\mu {\tilde \psi}~ \partial_\mu
{\hat{\phi}} - 4 \left(\tilde{m_t} + \frac{g_t}{\sqrt{2}} \tilde{H} 
\right)\frac{\hat{\phi}}{\vphi}{\overline{\tilde{\psi}}} {\tilde
\psi}\right] \noindent
\nonumber \\
+  \int d^4 x \left[ 3~ {\overline{\tilde{\psi}}}i
\gamma^\mu \partial_\mu {\tilde \psi}~ \frac{\hat{\phi}^2}{\vphi^2} + 
\frac{3 i}{\vphi^2}~ {\hat{\phi}}~ {\overline{\tilde{\psi}}}\gamma^\mu
{\tilde \psi}~ \partial_\mu {\hat{\phi}} - 6 \left(\tilde{m_t} 
+ \frac{g_t}{\sqrt{2}} \tilde{H}
\right)\frac{\hat{\phi}^2}{\vphi^2}{\overline{\tilde{\psi}}}
{\tilde \psi}\right].
\eea
In above $\psi = \left(\frac{f}{\vphi}\right)^{3/2} {\tilde{\psi}}$, 
~$H = \left(\frac{f}{\vphi}\right) {\tilde{H}}$~ and~ 
$m = \left(\frac{f}{\vphi}\right) {\tilde{m}}$. In the following we shall assume that all fields and 
parameters have been properly scaled so as to corresponds to the TeV scale and drop the
$\it{tilde}$ sign.

We use the same cut-off regularization technique with the 
UV cut-off $\L$ for the 
following vertex renormalization diagrams (see Figure 2a) to determine the renormalized $g_t$. 
To find the contribution of these diagrams to
$H{\overline{\psi}} \psi$ vertex we have to consider only those
terms in the loop integral that do not depend on the external momentum.
\newpage
\newpage
\begin{figure}[htb]
\begin{center}
\vspace*{4.5in}
      \relax\noindent\hskip -6.4in\relax{\includegraphics{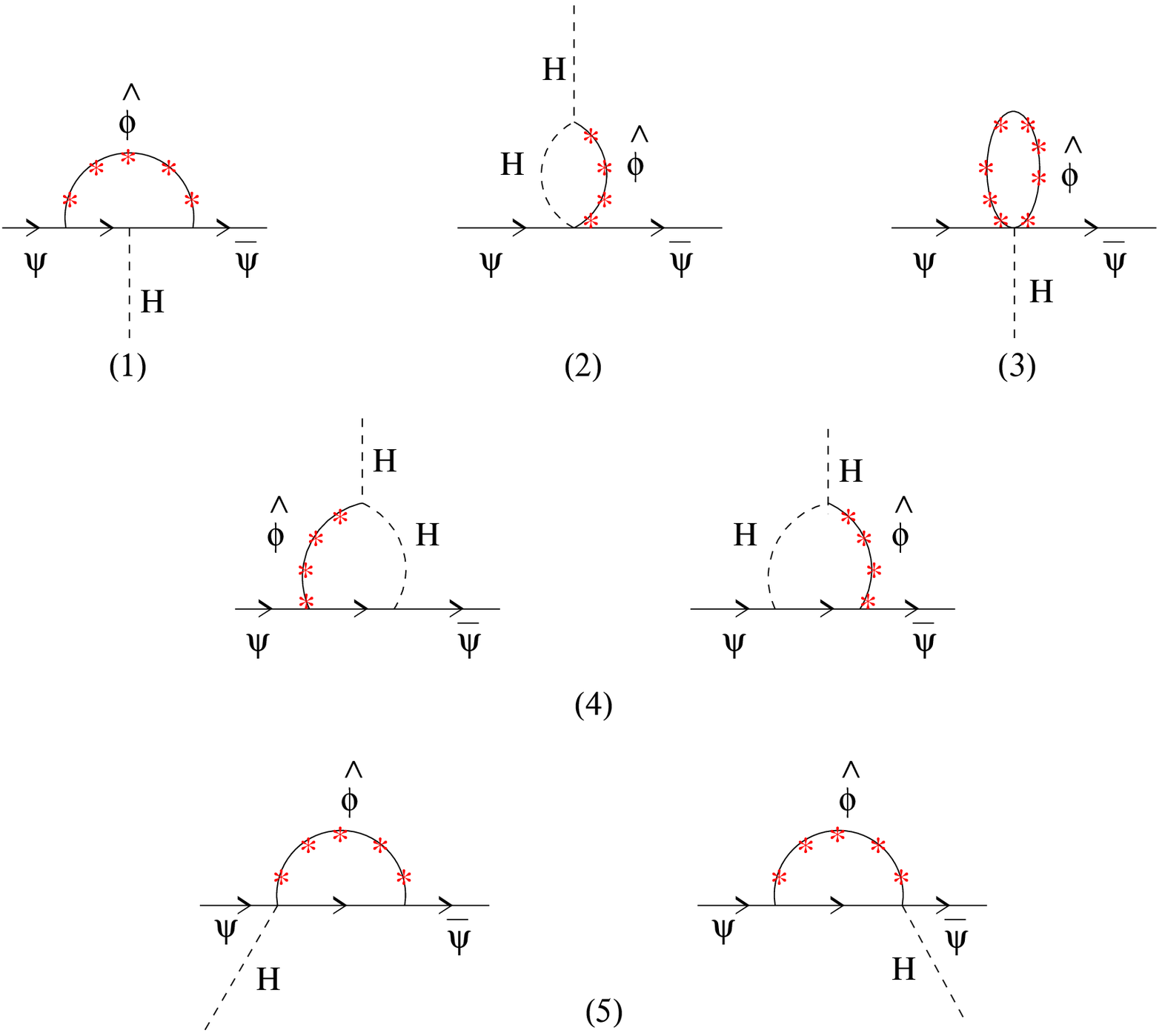}}
\end{center}
\end{figure}
\vspace*{0.2in}
\noindent { Figure 2a:} {\it {Feynman diagrams that give rise to radion contribution to
$H{\overline{\psi}} \psi$ vertex correction.}}

The reason being external momentum will give rise to derivative of
external fields and there are no such derivatives in the Yukawa term
$H{\overline{\psi}} \psi$. Considering only the external momentum
independent terms and retaining only the contributions of 
such terms that diverge with
the cut off $\L$ we get 

\bea
\Gamma_1 = -\left(\frac{g_t}{\sqrt{2}}\right) \frac{1}{16 \pi^2
\vphi^2}\left[\frac{9}{4} \L^2 - \frac{1}{4}(9 m_\phi^2 - 5
m_t^2)~ln{\frac{\L^2}{\mu^2}}\right], 
\eea
\bea
\Gamma_2 = \left(\frac{g_t}{\sqrt{2}}\right) \frac{1}{16 \pi^2
\vphi^2}\left[16~ m_h^2~ ln{\frac{\L^2}{\mu^2}}\right],
\eea
\bea
\Gamma_3 = -6 \left(\frac{g_t}{\sqrt{2}}\right) \frac{1}{16 \pi^2 
\vphi^2}\left[\L^2 -  m_\phi^2 ~ln{\frac{\L^2}{\mu^2}}\right], 
\eea
\bea
\Gamma_4 = - 12 \left(\frac{g_t}{\sqrt{2}}\right) \frac{1}{16 \pi^2   
\vphi^2}\left[m_h^2~ ln{\frac{\L^2}{\mu^2}}\right],
\eea
\bea
\Gamma_5 = - \left(\frac{g_t}{\sqrt{2}}\right) \frac{1}{16 \pi^2
\vphi^2}\left[- 12 \L^2 + (12 m_\phi^2 - 20 m_t^2) 
~ln{\frac{\L^2}{\mu^2}}\right],
\eea

\noindent where $\mu$ is the renormalization  scale. The wave function 
renormalization 
constant $Z_t$ of the top quark arise from the Feynman diagrams shown in Figure 2b

\vspace*{-0.8in}
\begin{figure}[htb]
\begin{center}
\vspace*{2.0in}  
      \relax\noindent\hskip -4.4in\relax{\includegraphics{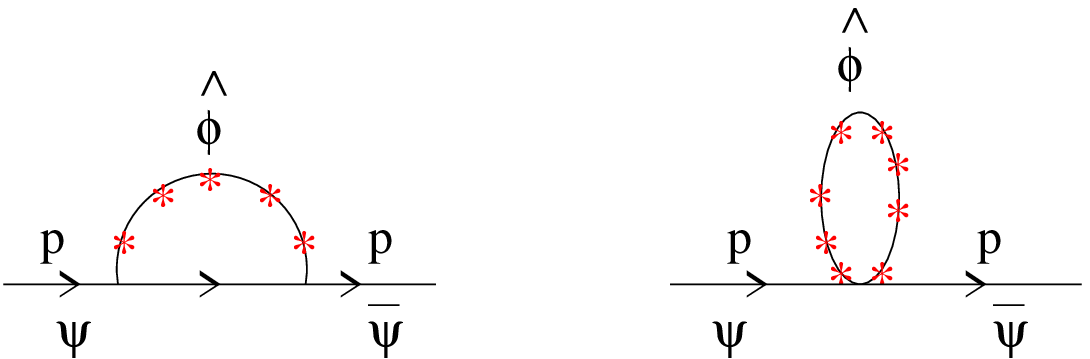}}
\end{center}
\end{figure}
\vspace*{-0.2in}
\noindent { Figure 2b:}
{\it { Feynman diagrams giving rise to $Z_t$.}}

\noindent Considering the terms proportional to 
$\slash{p}$ of Figure 2b, it can be shown that
\bea
Z_t =  1 +  \frac{1}{16 \pi^2 \vphi^2}\left[ \frac{39}{8} \L^2 - 
6 m_\phi^2 ~ln{\frac{\L^2}{\mu^2}} + \frac{13}{4} m_t^2
~ln{\frac{\L^2}{\mu^2}} \right].      
\eea

Following the above vertex and wave function renormalization constants,
it can be shown that the radion contribution $g_t(\mu )$ to the renormalized 
Yukawa coupling is given by
\bea
g_t(\mu) =  \frac{g_t}{16 \pi^2 \vphi^2}\left[ \frac{9}{8} \L^2 - 2
m_\phi^2~ln{\frac{\L^2}{\mu^2}} - \frac{31}{2}
m_t^2~ln{\frac{\L^2}{\mu^2}} -  \frac{9}{4}m_h^2~ln{\frac{\L^2}{\mu^2}} 
\right].
\eea

The complete beta function for $g_t$ in the presence of radion
fluctuations one finds as \cite{DasUma2,DasUma3}
\bea \label{eqn:hgtRad}
\beta(g_t(\mu)) = \beta_{SM}(g_t(\mu))+  
\frac{g_t}{16 \pi^2 \vphi^2}\left[ 4 m_\phi^2 + \frac{31}{2} g_t^2 v^2 + 9
\l v^2 \right],
\eea                           
where
\bea
\beta_{SM}(g_t(\mu)) = \frac{g_t}{16 \pi^2}
\left[\frac{9}{2} g_t^2 - 8 g_3^2 - \frac{9}{4} g_2^2 -
\frac{17}{12} g_1^2\right]
\eea
is the pure SM part \cite{BHL}.

\newpage
\subsection{Radion contribution to muon anomaly}
We now find the radion contribution to the muon anomaly.
The possible Feynman diagrams contributing to muon anomaly are shown in 
Figures  3(a,~b,~c,~d):
 
\vspace*{-1.0in}
\begin{figure}[htb]
\begin{center}
\vspace*{5.5in}
      \relax\noindent\hskip -5.4in\relax{\includegraphics{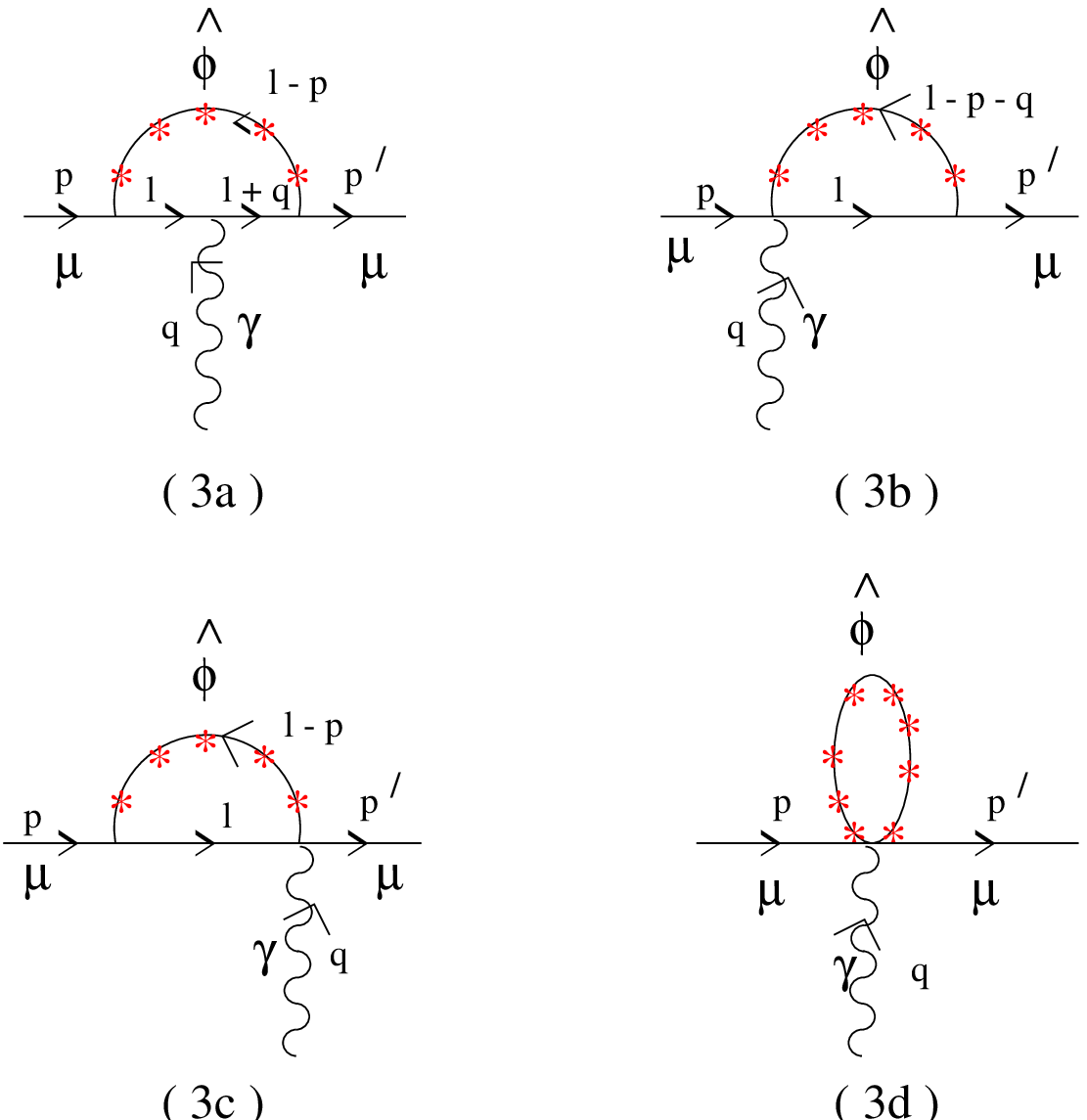}}
\end{center}
\end{figure}
\vspace*{-0.2in}
\noindent {Figures 3[a,b,c,d]}:
{{ \it Feynman diagrams contributing to muon  anomalous magnetic moment.}}

We regularize these diagrams by using the cut-off regularization 
technique with the ultra-violet cut-off as $\L$. A glimpse of the main results are presented 
below 
\bfl
For Figure 3(a):
\bea
- i e \Lambda_{1\mu} (p,q,p^\prime) = \frac{9 e}{2 \vphi^2} \int_0^1 
\int_0^1 x dx
dy \int \frac{d^4 l}{(2 \pi)^4} \frac{[\slash{l}_0 + 2 \slash{p^\prime} -
\frac{8}{3}m] [\slash{l}_0 + \slash{p^\prime} + m]\gamma_\mu[\slash{l}_0 +
\slash{p} + m][\slash{l}_0 + 2 \slash{p} - \frac{8}{3}m]}{[l^2 - R_1^2]^3} 
\nonumber\\
\eea
\efl

\noindent which gives
\vspace*{-0.25in}
\bfl
\bea
\Lambda_{1\mu} = \frac{9 i}{2 \vphi^2} \int_0^1  x dx \int_0^1 dy 
\left[\frac{2}{3}m_\mu (2 x
y - 1)\right] \left[\frac{i}{16 \pi^2} Log[\L^2/R_1^2] - \frac{3 i}{32 \pi^2}
\right] (p + p^\prime)_\mu \nonumber \\
\noindent 
+ \frac{9 i}{2 \vphi^2} \int_0^1 x dx \int_0^1 dy  
\left[\frac{2}{3} m_\mu^3 A\right] [- \frac{i}{32 \pi^2 R^2}]
(p + p^\prime)_\mu,  
\eea
where $l_0 = l - p x y - p^\prime x (1 - y)$, 
$R_1^2 = m_\mu^2 x^2 + m_\phi^2 (1 - x)$ 
 and $A = 2 x^3 y + \frac{4}{3} x^2 y - x^2 - \frac{8}{3}x y $.\\
\efl

\bfl
For Figure 3(b)
\bea
- i e \L_{2\mu} (p,q,p^\prime) =  - \frac{9 e}{2 \vphi^2}\int_0^1 dx \int 
\frac{d^4 l}{(2 \pi)^4} \frac{[\slash{l_{1}} + 2 \slash{p}^\prime - \frac{8}{3}
m][\slash{l_{1}} + \slash{p}^\prime + m ] \gamma_\mu}{[l^2 - R_2^2]^2} 
\nonumber \\
\noindent 
= - \frac{9 e}{2 \vphi^2}\int_0^1 dx \int \frac{d^4 l}{(2 \pi)^4} 
\frac{\left[l^2 + m^2 x^2 - 2 x m^2 + \frac{2}{3} x m - \frac{4}{3} m^2\right] 
\gamma_\mu} {[l^2 - R_2^2]},
\eea
\efl
where $R_2^2 = m_\mu^2 x^2 + m_\phi^2 (1 - x)$ 
and $\slash{l_{1}} = \slash{l} - x \slash{p}^\prime$.

\bfl
Similarly for Figure 3(c)
\bea
- i e \L_{3\mu} (p,q,p^\prime) =  - \frac{9 e}{2 \vphi^2}\int_0^1 dx \int 
\frac{d^4
l}{(2 \pi)^4} \frac{\gamma_\mu[ \slash{l_{2}} + \slash{p} + m]
[\slash{l_{2}} + 2\slash{p} - \frac{8}{3} m]}{[l^2 - R_3^2]^2}\nonumber \\
\noindent
= - \frac{9 e}{2 \vphi^2}\int_0^1 dx \int \frac{d^4 l}{(2 \pi)^4} 
\frac{\left[l^2 + m^2 x^2 -\frac{4}{3} m^2 x - \frac{4}{3} m^2\right] 
\gamma_\mu} {[l^2 - R_3^2]},
\eea
where $R_3^2 = m_\mu^2 x^2 + m_\phi^2 (1 - x) = R_2^2 = R_1^2 = R^2$ (say) and 
$\slash{l_{2}} = \slash{l} - x \slash{p}$. 

\efl

\bfl
Finally for  Figure 3(d), we find
\bea
- i e \L_{4\mu} (p,q,p^\prime) =  \frac{3 e}{\vphi^2}\gamma_\mu \int \frac{d^4
l}{(2 \pi)^4} \frac{1}{l^2 - m_\phi^2}, 
\eea
which  gives
\bea
\L_{4\mu} = \frac{3}{16 \pi^2 \vphi^2}\left[\L^2 -
m_\phi^2 Log\left(\frac{\L^2}{m_\phi^2}\right)\right] \gamma_\mu. 
\eea
\efl

\noindent It is clear from the above expressions of $\L_{2\mu}$,$\L_{3\mu}$ and 
$\L_{4\mu}$ that they are proportional to $\gamma_\mu$ and hence the
Feynman diagrams 3(b), 3(c) and 3(d) do not contribute to the muon anomalous magnetic
moment, but they do contribute in the vertex i.e. coupling constant renormalization.
On the other hand $\L_{1\mu}$ corresponding to the Figure 3(a) is seen to be
proportional to $(p + p^\prime)_\mu$ and 
contribute to $\delta a_\mu^{(new)} (=a_\mu^\phi)$. Using the Gordan's identity 
\bea
{\overline{u}}(p^\prime) \gamma_\mu u(p) = \frac{1}{2 m_\mu}{\overline{u}}
(p^\prime) \left[ (p + p^\prime)_\mu + i \sigma_{\mu\nu} q^\nu\right] u(p).
\eea

\noindent and the Dirac equation of motion we finally  get the 
radion contribution to the muon anomalous magnetic moment as \cite{MD}

\bea \label{eqn:muonanly}
a_\mu^\phi = \frac{36 m_\mu^2}{96 \pi^2 \vphi^2} \int_0^1 x dx \int_0^1 
dy~ (2 x y - 1)\left[ Log\left(\frac{\L^2}{R^2}\right) - \frac{3}{2}\right] 
\nonumber \\
\noindent
- \frac{36 m_\mu^4}{192 \pi^2 \vphi^2} \int_0^1 x dx \int_0^1 dy~ 
\frac{\left[2 x^3 y + \frac{4}{3}x^2 y  - x^2 - \frac{8}{3} x y \right]}{R^2},
\eea

\noindent where $R^2 = m_\mu^2 x^2 + m_\phi^2 (1 - x)$, $m_\mu$ the muon mass. At this point it is 
worthwhile to note that the radion mediated muon anomaly 
is free from power like divergence unlike the Kaluza-Klein graviton 
contribution to the oblique electroweak parameters S, T and U which 
is plagued by uncalculable powerlike divergences \cite{DR,HMZ}.
Now $\L$ in Eq. (\ref{eqn:muonanly}) is
the ultraviolet cut-off of the theory and from a naive 
dimensional analysis it follows that $\L$ is equal to $4 \pi \vphi$ \cite{GM}. The UV  limit $\L \sim \langle \phi \rangle \rightarrow \infty$ corresponds to the radion 
and SM  decoupling limit.


\section{Numerical Analysis}
There are phenomenological limits
on the $m_\phi - \vphi$ parameter space. From this it follows that the lower
bound on $\vphi$ can range from
about the electroweak symmetry breaking scale to about a TeV, while
$m_\phi$ can in principle be even lighter than $m_W$ or heavier than the 
top quark. We seperate our analysis 
in two main parts. First, we compare our radion corrected muon 
anomaly $a^\phi_\mu$ 
with the deviation $\delta a^{new}_\mu$ (using the BNL recent result) 
and obtain 
constraints in the $m_\phi - \la \phi \ra$ plane. Second, we use the modified beta functions $\beta(\l)$ 
and $\beta(g_t)$ together with the anomaly constrained $m_\phi$ 
and $\la \phi \ra$ values to obtain lower bound on higgs mass $m_h$. 
Finally, we compare our result with the LEP2 direct bound on $m_h$. 

\subsection{Anomaly constraints in $m_\phi - \vphi$ plane}  
The radion contribution to the muon anomaly ( 
Eqn.(\ref{eqn:muonanly})) in the limit $\L \gg m_\phi \gg m_\mu$ takes the 
form
\bea
a_\mu^{(\phi)} = \frac{36 m_\mu^2}{96 \pi^2 \vphi^2} \left[0.12 
- 0.17~ Log\left(\frac{16 \pi^2 \vphi^2}{m_\phi^2} \right)
-0.26 ~\frac{m_\mu^2}{m_\phi^2}\right].
\eea 

From Eqs. (\ref{eqn:muonanlyBNL}) and (\ref{eqn:muonanlySM}) we see that 
the experimental result differs from the theoretical(SM) prediction by 
\bea
\delta a_\mu^{new} = a_\mu^{(expt)} - a_\mu^{(SM)} 
= (28 \pm 10.5) \times 10^{-10}
\eea
which is about $2.6~\sigma$.
The ultimate precision of the BNL collaboration is to
reduce the error down to $4.0 \times 10^{-10}$. We will consider the BNL recent result for our analysis 
and make some comments regarding the ultimate precission measurement.

\vspace*{-0.25in}
\begin{figure}[htb]
\begin{center}
\vspace*{3.5in}
      \relax\noindent\hskip -4.4in\relax{\includegraphics{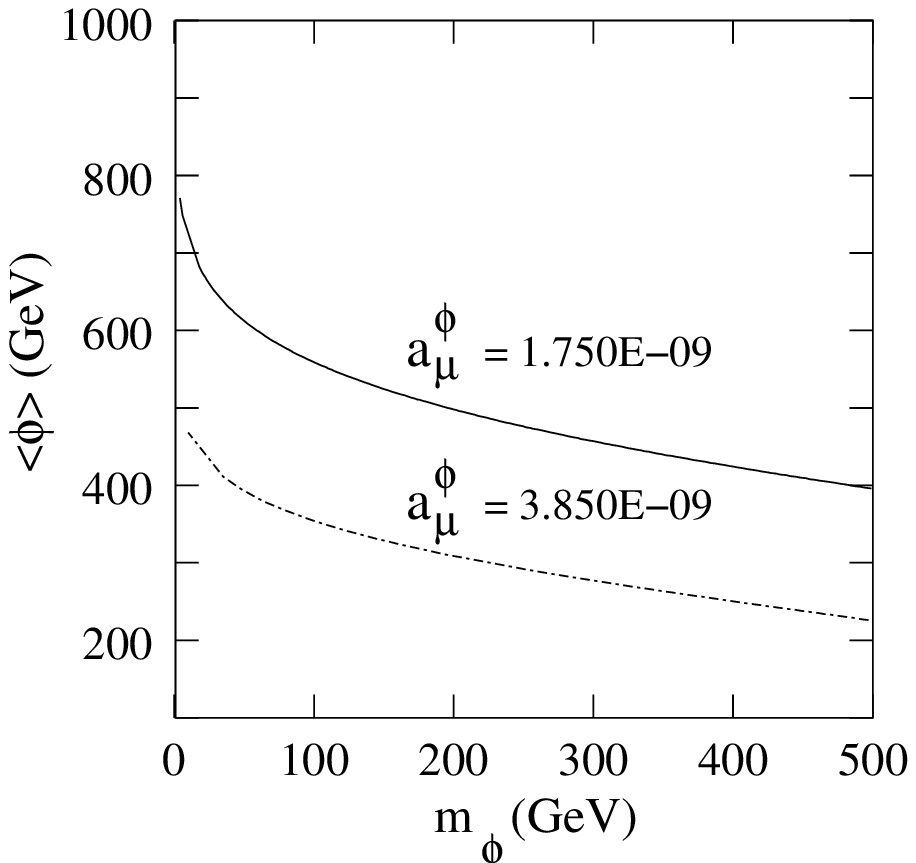}}
\end{center}
\end{figure}
\vspace*{-0.35in}
\noindent {Figure 4}.
{{ \it Muon anomaly constraints on 
$\mphi$ and $\vphi$. For any curve the allowed region lies at and above 
the curve.}}

In Figure 4 we draw the contour plots in $m_\phi - \vphi$ plane 
corresponding to of $a_\mu^{\phi} [= \delta a_\mu^{new}]= 1.75\times 10^{-9}$ and $3.85\times 10^{-9}$.
The following things are to be noted:
\begin{itemize}
\item For a given $m_\phi$ the lower bound on $\vphi$ increases with the decrease 
in $a_\mu^{\phi}$. As an example for $m_\phi = 100$ GeV when 
$a_\mu^\phi$ varies from $3.850 \times 10^{-9}$ to
$1.750 \times 10^{-9}$, $\vphi$ changes from $354$ GeV to $559$ GeV.
\item For a given $a_\mu^\phi$ the lower bound on $\vphi$ decreases with the increase 
in $m_\phi$.
\item For any curve the region at and above the curve is allowed. 

\item Elaborating Figure 4, we find that the BNL ultimate(projected) precission measurement  
suggests the lower bound on $\vphi$ larger than $400$ GeV for a
heavy radion (for $\mphi = 500$ GeV), while $\vphi> 1000$ GeV for a lighter one i.e. 
$\mphi = 100$ GeV.
\end{itemize}

\subsection{Lower bound on higgs mass $m_h$}
To obtain bound on $m_h$ rewrite $\beta(\lambda)$ and 
$\beta(g_t)$ as 

\vspace*{-0.25in}
\bea
\beta(\lambda) =  \frac{d \lambda}{d t} = \mu \frac{d \lambda}{d \mu} 
=\frac{1}{8 \pi^2} \left [ 9
\lambda^2 + \frac{402 \lambda^2 v^2}{\vphi^2} + \frac{144 \lambda^2
v^4}{\vphi^4} + \frac{5 \lambda m^2_{\phi}}{\vphi^2} + \lambda \left (6
~g^2_t - \frac{9}{2}~ g_2^2 - \frac{3}{2}~ g_1^2 \right ) - 6~ g^4_t
\right ] \noindent
\nonumber \\
+~\frac{1}{8 \pi^2} \left [\frac{3}{4} \left( g_2^4 + \frac{1}{2}
{(g_2^2 + g_1^2)^2} \right )\right ] ~(14) \nonumber 
\eea
and
\bea
\beta(g_t) = \frac{d g_t}{d t} = \beta_{SM}(g_t)+
\frac{g_t}{16 \pi^2 \vphi^2}\left[ 4 m_\phi^2 + \frac{31}{2} g_t^2 v^2 + 9
\l v^2 \right],~(29) \nonumber 
\eea
where $$\beta_{SM}(g_t) = \frac{g_t}{16 \pi^2}
\left[\frac{9}{2} g_t^2 - 8 g_3^2 - \frac{9}{4} g_2^2 -
\frac{17}{12} g_1^2\right].$$

\noindent In above $v (= 247$ GeV) is the electro-weak vev, $g_2$ 
and $g_1$ are the $SU(2)_L$ and $U(1)_Y$ coupling
constants and $t = log(\frac{\mu}{\mu_0})$ with $\mu$, the 
renormalization scale and $\mu_0$, a reference scale (in our case 
is chosen as $m_Z$). Note that the terms which arise as corrections due to radion in $\beta(\lambda)$ 
and  $\beta(g_t)$ goes in powers of $1/{\la \phi \ra}$. Rest are the SM terms.

We combine the anomaly constrained $m_\phi$ and $\vphi$ 
with the above modified $\beta(\lambda)$ and $\beta(g_t)$ to obtain 
bound on $m_h$. For this we consider the following steps:

 1. In Figure 4 we have seen that for a given $a_\mu^\phi$ the
lower bound on $\vphi$ decreases with the increase of $m_\phi$. Although for a given curve the region at 
and above the curve 
is allowed, we will choose the points on the curves.

 2. Now we find the higgs quartic 
coupling $\lambda (\mu = 115)$ (GeV) by
solving the beta function $\beta(\lambda)$ ( Eq.~(\ref{eqn:hdaRad}))
for the following two initial value (required while running them from top to bottom):\\ 
(i) $\lambda(\L = 4 \pi \vphi) = 3.54491$, 
non-perturbative and \\
(ii) $\lambda(\L= 4 \pi \vphi) = 0.313$ i.e. perturbative.

Note that in Eqs. (\ref{eqn:hdaRad}) and (\ref{eqn:hgtRad}) the coupling constants $g_1$, 
$g_2$ and $g_3$ are renormalized coupling constants. Their relevant beta functions(to one loop order)
are given by
\bea
\beta(g_1) = \frac{41}{96 \pi^2}~ g_1^3,
\eea
\vspace*{-0.25in}
\bea
\beta(g_2) = - \frac{19}{96 \pi^2}~ g_2^3,
\eea
and
\bea
\beta(g_3) = - \frac{7}{16 \pi^2}~ g_3^3.
\eea
   
\noindent In solving the above renormalization group(RG) equations, we use the  
following inputs 
$g_t(\mu=m_Z) = \frac{{\sqrt{2}} m_t}{v} = 1.001$, $g_2(m_Z) = \frac{e}{Sin\theta_w}
= 0.644$ and $g_1(m_Z) = \frac{e}{Cos\theta_w} = 0.356$. 
Using the above RG equations we next run all the coupling constants. First we allow them to run 
from $\mu = m_Z$
to $\mu = \Lambda$ and note their values at $\mu = \Lambda(= 4 \pi \vphi)$. 
Then we run them from top to bottom i.e. from  $\mu = \Lambda$ to $\mu = 115$ GeV corresponding to  
$\lambda(\L = 4 \pi \vphi) = 3.54491$ and $\lambda(\L= 4 \pi \vphi) = 0.313$ and note 
$\lambda(\mu = 115$ GeV) accordingly.
Plots showing $\lambda (\mu = 115$ GeV) as a function of $\vphi$, the UV cut-off $\L$ 
($= 4 \pi \vphi)$ are shown in Figures 5[a,~b] corresponding to non-perturbative and perturbative 
initial conditions.

3. In Figure 5a (the non-perturbative case) we have several distinct 
lines corresponding to $\lambda(\mu = 115 GeV)$ vs $\la \phi \ra$ (GeV) plots 
for different $m_\phi$ values.
Now for a particular $m_\phi$, $\la \phi \ra$ (which is consistent with the muon anomaly
constraint (see Figure 4)) is 
chosen from Figure 5a and the corresponding $\lambda(\mu = 115$ GeV) is noted.
A similar kind of analysis is also done for the perturbative case and accordingly Figure 5b is 
obtained.

4. After finding $\lambda(\mu = 115$ GeV) corresponding to 
a given $m_\phi$ and $\la \phi \ra$ (which are consistent with the muon anomaly 
constraint), we can convert it to the higgs
mass $m_h$ by using the relation $$m_h(\mu) = \sqrt{2 \lambda(\mu)~ v^2}$$ respectively for 
the perturbative and non-perturbative cases. 

\vspace*{-0.5in}
\newpage
\begin{figure}
\subfigure[]{
\label{PictureOneLabel}
\hspace*{-0.7 in}
\begin{minipage}[b]{0.5\textwidth}
\centering
\includegraphics[width=\textwidth]{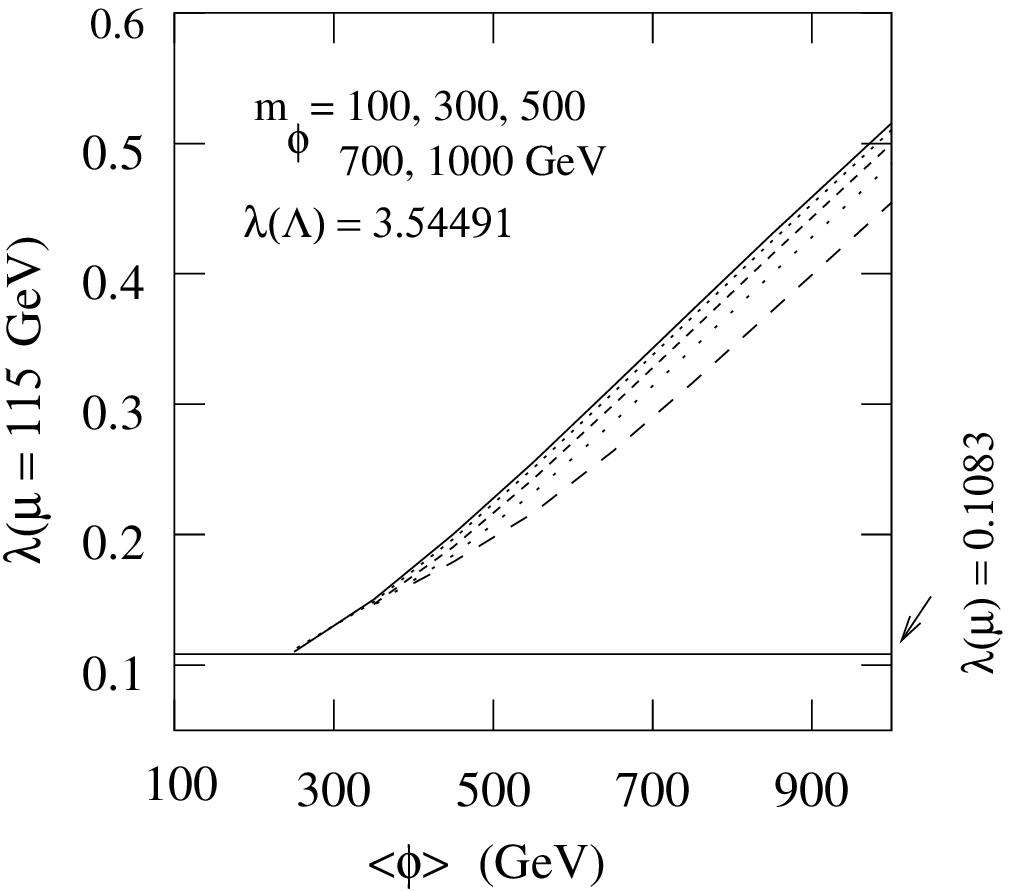}
\end{minipage}}
\subfigure[]{
\label{PictureTwoLabel}
\hspace*{0.3in}
\begin{minipage}[b]{0.5\textwidth}
\centering
\includegraphics[width=\textwidth]{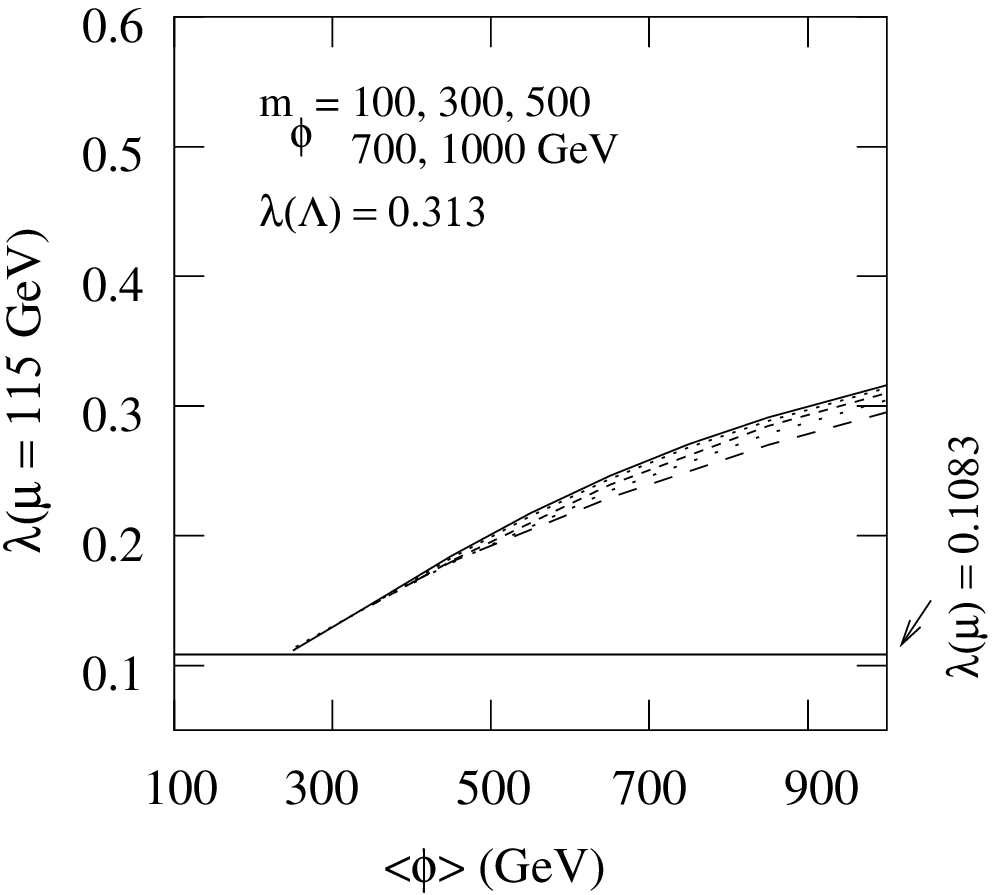}
\end{minipage}}
\end{figure}
\vspace*{-0.45in}
\noindent {Figures 5[a,~b]}.
{{ \it Plots showing $\l(\mu = 115 GeV)$ as a function of $\vphi$ with 
$\l(\L) = 3.54491$ and $=0.313$ for different $m_\phi$ values.}}
\vspace*{0.15in}

 5. We next plot the higgs mass $m_h$ as a function of $m_\phi$ 
for $a_\mu^\phi = 1.750 \times 10^{-9}$ and $ 3.850 \times 10^{-9}$ respectively for $\l(\L) = 3.5449$ 
and $\l(\L) = 0.313$ in either case and
they are shown in
Figures 5c and 5d. The horizontal line (in both figures) corresponds to the LEP2
lower bound on $m_h$ which is about $115$ GeV (See \cite{Junk} for LEP2 direct search of higgs boson). 

 6. In Figure 5c the region {\bf A} is allowed both by the
LEP2 direct search and the muon anomaly 
$\delta a_\mu^{(new)} = a_\mu^\phi = 1.750 \times 10^{-9}$ and it gives a lower 
bound on $m_h$ which varies from $190$ GeV to $142$ GeV for $m_\phi$ ranges
from  $1$ GeV to $500$ GeV. 
It is to be noted that the lower bound on $m_h$ corresponding to
$a_\mu^\phi = 1.750 \times 10^{-9}$ is all throughout greater than
the LEP2 bound. 
Similarly in Figure 5d, corresponding to
$a_\mu^\phi = 1.750 \times 10^{-9}$ we see that the lower bound on $m_h$ 
varies from $170$ GeV to $141$ GeV for $m_\phi$ ranges from  $1$ GeV
to $500$ GeV.

 7. The region {\bf B} in Figures 6[c,d] is 
allowed both by the direct LEP2 search and 
$a_\mu^\phi = 3.850 \times 10^{-9}$, but forbidden by
$a_\mu^\phi = 1.750 \times 10^{-9}$. Interesting bound on $m_h$ follows from this region. 
We find a lower bound of about $115$ GeV on 
$m_h$ which is compatible with the LEP2 bound. We also
obtain the upper bounds on $m_h$ depending on whether $\l(\L)$ is
non-perturbative or perturbative. They are about $142$ GeV and $141$ 
GeV corresponding to $m_\phi = 500$ GeV respectively for  $\l(\L) = 3.5449$   
and $\l(\L) = 0.313$. 

\vspace*{-0.5in}
\newpage
\begin{figure} 
\subfigure[]{
\label{PictureThreeLabel}
\hspace*{-0.7 in}
\begin{minipage}[b]{0.5\textwidth}
\centering
\includegraphics[width=\textwidth]{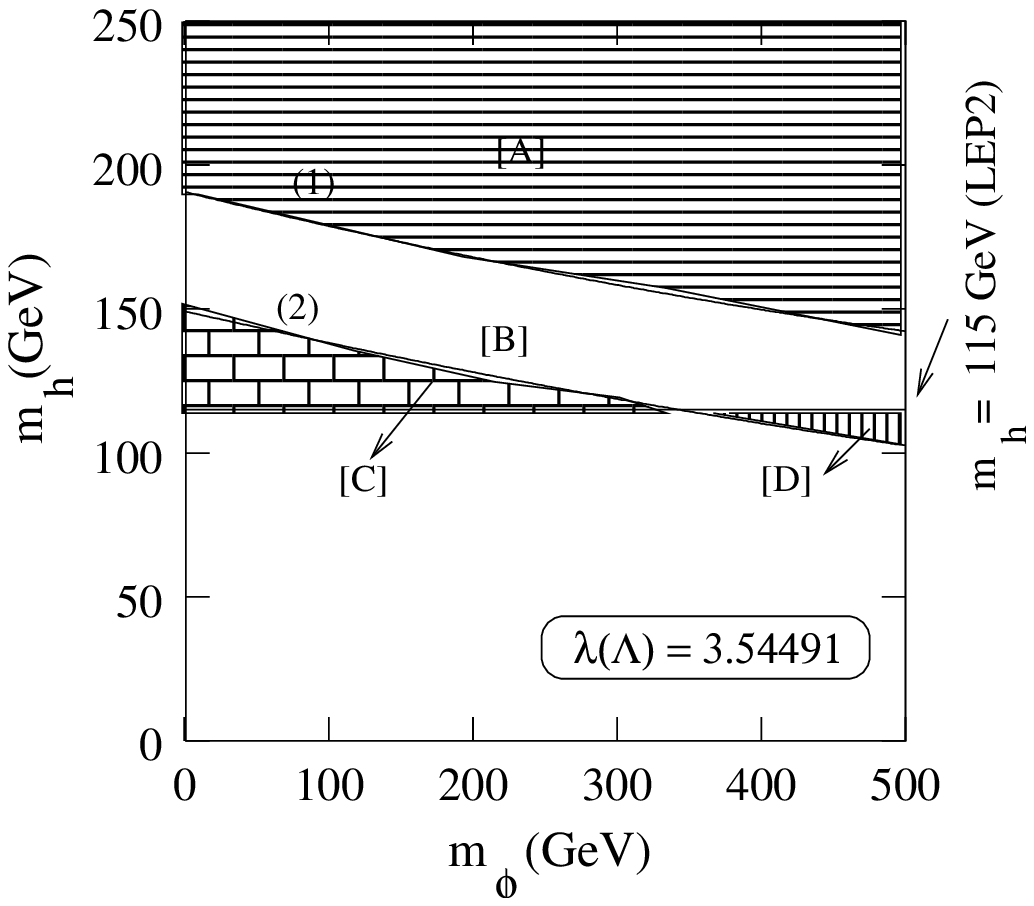}
\end{minipage}}
\subfigure[]{
\label{PictureFourLabel}
\hspace*{0.3in}
\begin{minipage}[b]{0.5\textwidth}
\centering   
\includegraphics[width=\textwidth]{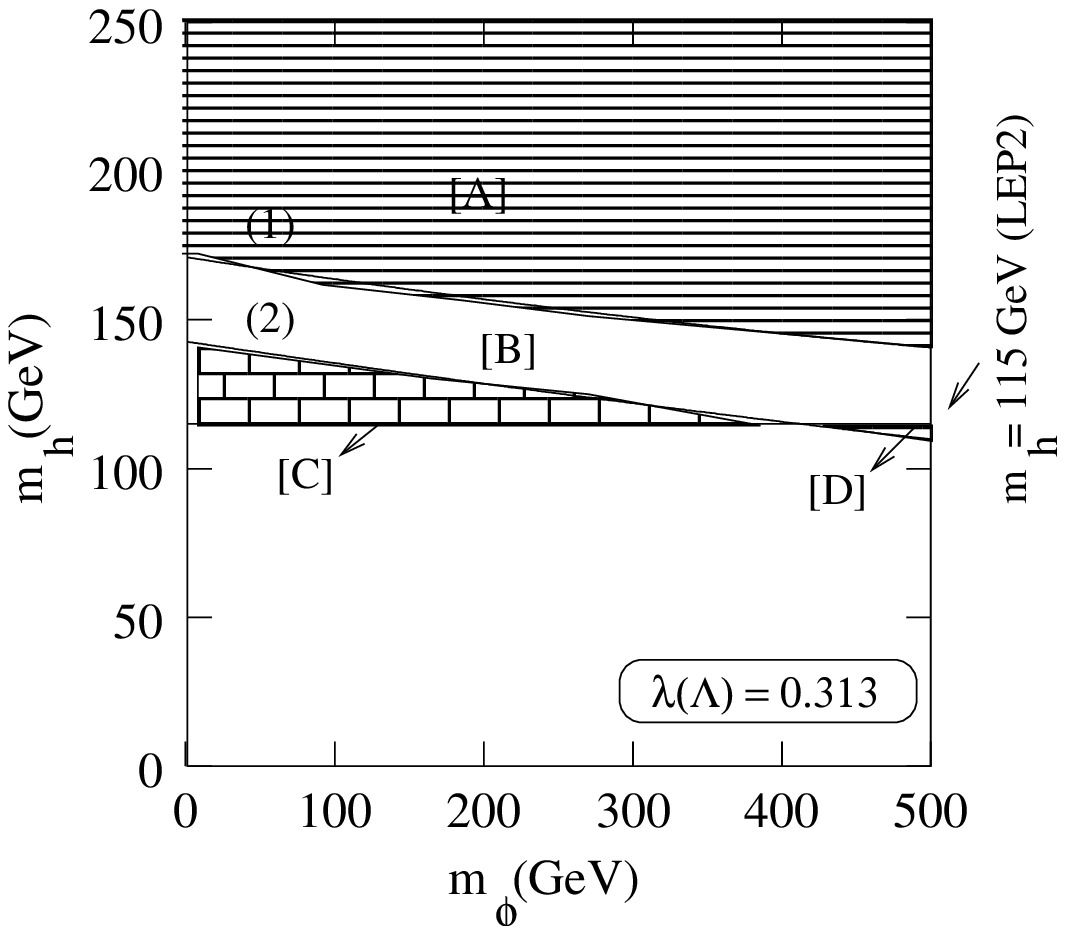}
\end{minipage}} 
\end{figure}
\vspace*{-0.4in}
\noindent {Figures 5[c,d]}.
{{ \it The $m_\phi$ vs $m_h$  plots corresponding to
$\l(\L) = 3.54491$ and $=0.313$. The upper and the lower curve
corresponds to (1) $a_\mu^\phi = 1.750 \times 10^{-9}$ and
(2) $a_\mu^\phi = 3.850 \times 10^{-9}$.}}


\vspace*{0.15in}

 8. The region {\bf C} is allowed by the direct LEP2 
search but disallowed by $\delta a_\mu^{(new)} = a_\mu^\phi$.
In addition of finding a lower bound on $m_h$ which is compatible with the
LEP2 bound, we also obtain an upper bound of about $149$ GeV for the
non-perturbative case and $143$ GeV for the perturbative case
corresponding to a very light radion (say $m_\phi = 1 \sim 2$ GeV).
{\it Most importantly the bound on $m_h$ corresponding to
$a_\mu^\phi = 3.850 \times 10^{-9}$ is greater than the LEP2 bound
(which is about $115$ GeV) if $m_\phi < 342$ GeV for $\l(\L) = 3.54491$  
and is $< 412$ GeV for $\l(\L) = e (=0.313)$. Translating this to $\vphi$
one finds the lower bounds on $\vphi$ as $266$ GeV and $247$ GeV
corresponding to $a_\mu^\phi = 3.850 \times 10^{-9}$}. 

 9. Finally the region {\bf D} in Figures 5c and 5d which is ruled out by direct LEP2
search, allowed by $\delta a_\mu = 3.850 \times 10^{-9}$. Since it is forbidden by LEP2 data, we 
do not consider this region any further. 

\section {Summary and Conclusion:}
The model of warped spatial dimension (the Randall-Sundrum model) with a light stabilized radion 
posseses several interesting phenomenological features. We explore one such feature in the present work.
We calculate the radion mediated muon anomaly $a_\mu^{\phi}$ and use the BNL muon anomaly data 
to constrain $m_\phi$ and $\vphi$, the two free parameters of this model. The beta 
functions $\beta(\l)$, $\beta(g_t)$ for
the higgs quartic coupling $\l$ and the higgs-top Yukawa coupling 
$g_t$ gets modified
in the  presence of radion and we determine these modified functions. 
Using these modified beta functions and the anomaly 
constrained $m_\phi$ and $\vphi$ we obtain  lower bound on higgs 
mass $m_h$.  For $a_\mu^\phi = 3.850 \times 10^{-9}$, we find that 
the bound on $m_h$ is greater than the LEP2 bound 
if $m_\phi < 342$ GeV for $\l(\L) = 3.54491$ and is $< 412$ GeV for
$\l(\L) = 0.313$. Translating this to $\vphi$ one finds lower bound of  
$266$ GeV and $247$ GeV for $\l(\L) = 3.54491$ and $\l(\L) = 0.313$, respectively.
The bound on $m_h$ corresponding to
$a_\mu^{\phi} = 1.750 \times 10^{-9}$ is found to be greater than the 
LEP2 bound
for a wide range of $m_\phi$ both for perturbative and non-perturbative
values of $\l(\L)$.

\bfl
{\Large \bf {Acknowledgement:} }
\efl
I would like to thank late  Prof. Uma Mahanta of HRI who taught me the Randall-Sundrum 
model and to Prof. Biswarup Mukhopadhyaya of HRI 
for his useful suggestions and comments while reading the manuscript. Special thanks to Prof. Saurabh 
Rindani for providing me a nice stay at the PRL where this work is finally 
completed.


\vspace*{-0.15in}

\begin{thebibliography}{92}


\bibitem{ADD1} N. Arkani-Hamed, S. Dimopoulos and G. Dvali, {\it Phys. Lett.}
{\bf B429} , 263 (1998).

\bibitem{ADD2} I. Antoniadis, N. Arkani-Hamed and G. Dvali, {\it Phys.
Lett.} {\bf B463}, 257 (1998).

\bibitem{RS1} L. Randall and R. Sundrum, {\it Phys. Rev. Lett.} {\bf 83}, 3370 (1999). 

\bibitem{RS2} L. Randall and R. Sundrum, {\it Phys.Rev.Lett.} {\bf 83}, 4690 (1999).

\bibitem{GMPK} M. L. Graesser, {\it Phys. Rev.} {\bf D 61}, 074019 (2000). 

\bibitem{MR} U. Mahanta and S. Rakshit, {\it Phys. Lett.} {\bf B480}, 176 (2000).

\bibitem{PS} S. C. Park and H. S. Song, {\it Phys. Lett.} {\bf B506}, 99 (2001).

\bibitem{KimKim} C. S. Kim, J. D. Kim and J. Song, 
{\it Phys. Lett.} {\bf B511}, 251 (2001).

\bibitem{GRW} G. F. Giudice, R. Rattazzi and J. D. Wells, {\it Nucl. Phys.} 
{\bf B595}, 250 (2001).

\bibitem{GW} W. D. Goldberger and M. B.Wise, {\it Phys. Lett.} 
{\bf B475} 275-279 (2000). 

\bibitem{GR} W. D. Goldberger and I. Z. Rothstein, 
{\it Phys. Lett.} {\bf B491} 339 (2000).

\bibitem{GW1} W. D. Goldberger and M. B.Wise, {\it Phys. Rev. Lett.} {\bf 83}
4922 (1999).

\bibitem{GW2} W. D. Goldberger and M. B.Wise, {\it Phys. Rev.} {\bf D 60}
107505 (1999).

\bibitem{Cheung} K. Cheung, {\it Phys. Rev.} {\bf D63} 056007 (2001).

\bibitem{Csaki} C. Csaki, M. Graesser, L. Randall and J. Terning,
{\it Phys. Rev.} {\bf D62} 045015 (2000).

\bibitem{UmaDat} U. Mahanta and A. Datta, 
{\it Phys. Lett.} {\bf B483} 196 (2000).

\bibitem{BKL}S. Bae, P. Ko, H. Lee and J. Lee, {\it Phys. Lett.} {\bf B487} 299 (2000).

\bibitem{DM1} P. K. Das and U. Mahanta, {\it Phys. Lett.} {\bf 528} 253 (2002).

\bibitem{DM2} P. K. Das and U. Mahanta, {\it Mod. Phys. Lett.} {\bf A}, 127 
(2004).

\bibitem{CDHY} M. Chaichian, A. Datta, K.Huitu and Z. Yu,    
{\it Phys. Lett.} {\bf B524} 161 (2002).

\bibitem{E821ex1} G. W. Bennet et. al. [Muon g-2 collaboration],
hep-ex/0208001.
\bibitem{E821ex2} H. N. Brown et. al. {\it Phys. Rev. Lett.} {\bf 86}, 2227
(2001).

\bibitem{Hert} D. H. Hertzog, hep-ex/0202024.

\bibitem{KN} M. Knecht and A. Nyffeler, {\it Phys. Rev.} {\bf D65} 073034 (2002).

\bibitem{KNPR} M. Knecht, A. Nyffeler, M. Perrotet and E. de Rafael,
{\it Phys. Rev. Lett.} {\bf 88}, 071802 (2002).

\bibitem{HayKin} M. Hayakawa and T. Kinoshita, hep-ph/0112102.

\bibitem{BCM} I. Blokland, A. Czarnecki and K. Melnikov, {\it Phys. Rev. Lett.}
{\bf 88}, 071803 (2002).

\bibitem{CzarMar}A. Czarnecki and W.J. Marciano, {\it Phys. Rev.} {\bf D64}, 013014 (2001).

\bibitem{Lane} K. Lane, hep-ph/0102131.

\bibitem{EKRW} L. Everett, G. Kane, S. Rigolin and L. Wang, {\it Phys. Rev. Lett.} {\bf 86}, 3484 (2001).

\bibitem{FenMat} J.L. Feng and K.T. Matchev, {\it Phys. Rev. Lett.} {\bf 86}, 
3480 (2001). 

\bibitem{BalGon} E.A. Baltz and P. Gondolo, {\it Phys. Rev. Lett.} {\bf 86}, 
5004 (2001).

\bibitem{UtpalNath} U. Chattopdhyay and P. Nath, {\it Phys. Rev. Lett.} {\bf 86}, 5854 (2001).

\bibitem{Mahanta} U. Mahanta, {\it Phys. Lett.} {\bf B515}, 111 (2001).

\bibitem{CCD} D. Chakraverty, D. Choudhury 
and A. Datta,  {\it Phys. Lett.} {\bf B506}, 103 (2001).

\bibitem{CMR} D. Choudhury, B. Mukhopadhyaya
  and S. Rakshit, {\it Phys. Lett.} {\bf B507}, 219 (2001). 


\bibitem{BarMai} R. Barbieri and L. Maiani,
{\it Phys. Lett.} {\bf B117 }, 203 (1982).

\bibitem{KKS} D.A. Kosower, L.M. Krauss and N. Sakai,
{\it Phys. Lett.} {\bf B133}, 305 (1983).

\bibitem{ACN} R. Arnowitt, A.H. Chamseddine and P. Nath,
{\it Z. Physik} {\bf C26},407 (1984).

\bibitem{LNW} J.L. Lopez, D.V. Nanopoulos and X. Wang,
{\it Phys. Rev.} {\bf D49 }, 366 (1994).

\bibitem{AEW} C. Arzt, M.B. Einhorn and J. Wudka,
{\it Phys. Rev.} {\bf D49 }, 1370 (1994).

\bibitem{ChatNath} U. Chattopadhyay and P. Nath,
{\it Phys. Rev.} {\bf D53 }, 1648 (1996).

\bibitem{CGW} M. Carena, G.F. Giudice and C.E. Wagner,
{\it Phys. Lett.} {\bf B390}, 234 (1997).

\bibitem{NathYama} P. Nath and M. Yamaguchi,
{\it Phys. Rev.} {\bf D60 }, 116006 (1999).

\bibitem{CGV} R. Casadio, A. Gruppuso and G. Venturi,
{\it Phys. Lett.} {\bf B495 }, 378 (2000).

\bibitem{DHR} H. Davoudiasl, J.L. Hewett and T.G. Rizzo,
{\it Phys. Lett.} {\bf B493 }, 135 (2000).

\bibitem{MarCzar} A. Czarnecki and W.J. Marciano, hep-ph/0010194 and
  references therein.


\bibitem{Junk} T. Junk, The LEP Higgs Working Group at LEP Fest October 
10th, 2002.
                                                                                    
\bibitem{DRR} P. K. Das, S. K. Rai and S. Raychaudhuri, {\it Phys. Lett.} {\bf B618}, 221 (2005).

\bibitem{CDJ} J. C. Collins, A. Duncan and S. D. Joglekar, {\it Phys. Rev.}
{\bf D 16}, 438 (1977).

\bibitem{DasUma1} P. K. Das and Uma Mahanta, {\it Phys.Lett.} {\bf B520}, 
307 (2001).

\bibitem{DasUma2} P. K. Das and U. Mahanta, hep-ph/011030.

\bibitem{DasUma3} P. K. Das and U. Mahanta, {\it Int.J.Mod.Phys.} {\bf A20}, 1089 (2005).

\bibitem{BHL} W. Bardeen, C. Hill and M. Lindner, {\it Phys. Rev.} {\bf D41}, 
1647 (1990).

\bibitem{MD}P. K. Das and Uma Mahanta, {\it Nucl. Phys.} {\bf B644}, 
395-400 (2002).

\bibitem{DR} P. K. Das and S. Raychaudhuri, hep-ph/9908205.

\bibitem{HMZ} T. Han,
D. Marfatia and R. Zhang, {\it Phys. Rev.} {\bf D62}, 125018 (2000).

\bibitem{GM} H. Georgi and A. Manohar, {\it Nucl. Phys.} {\bf B234}, 
189 (1984).


\end{thebibliography}
\end{document}